\documentclass[twocolumn,twoside,slac]{revtex4}
\usepackage{graphicx}
\usepackage{fancyhdr}
\usepackage{epsfig}
\usepackage{axodraw}
\usepackage{amsfonts}
\usepackage{amssymb}
\begin{document}

\title{Challenges in Moving the LEP Higgs Statistics to the LHC}

%

\author{K.S. Cranmer, B. Mellado, W. Quayle, Sau Lan Wu}
\affiliation{University of Wisconsin-Madison, Madison, WI 53706, USA}

\begin{abstract}
We examine computational, conceptual, and philosophical issues in
moving the statistical techniques used in the LEP Higgs working group
to the LHC.
\end{abstract}

\maketitle

\thispagestyle{fancy}

\section{Introduction}\label{S:Intro}
Higgs searches at LEP were based on marginal signal
expectations and small background uncertainties.  In contrast, Higgs
searches at the LHC are based on strong signal expectations and
relatively large background uncertainties. Based on our experience
with the LEP Higgs search, our group tried to move the tools we had
developed at LEP to the LHC environment. In particular, our
calculation of confidence levels was based on an analytic computation
with the Fast Fourier Transform and the log-likelihood ratio as a test
statistic (and systematic errors based on the Cousins-Highland
approach). We encountered three types of problems when calculating
ATLAS' combined sensitivity to the Standard Model Higgs Boson:
problems associated with large numbers of expected events, problems
arising from very high significance levels, and problems related to
the incorporation of systematic errors.

Previously, it was shown that the migration of the statistical
techniques that were used in the LEP Higgs Working Group to the LHC
environment is not as straightforward as one might na\"{i}vely
expect~\cite{UW-confidence-VBF}. After a brief overview in Section~\ref{S:Formalism},
those difficulties and their ultimate solution are discussed in
Section~\ref{S:Numerical}. Our group has developed two independent software solutions
(both in C++; both with {\tt FORTRAN} bindings; one {\tt ROOT} based
and the other standalone) which can be found at:

\centerline{\tt http://wisconsin.cern.ch/software}

In Section~\ref{S:Systematics} we discuss the incorporation of
systematic errors and compare a few different strategies. In
Section~\ref{S:Luminosity} we present and discuss the discovery
luminosity (the luminosity expected to be required for
discovery). Lastly, in Section~~\ref{S:Power} we discuss the
statistical notion of {\it power} (which is related to the probability
of Type II error (the probability we do reject the
``signal-plus-background hypothesis" when it is true).

\section{The Formalism}\label{S:Formalism}
Our starting point for this note is a brief review of the techniques
that were used at LEP. We refer the interested reader
to~\cite{Kendall} for an introduction to the fundamentals,
to~\cite{Read:1997} for why the likelihood ratio has been chosen as a
test statistic, to~\cite{Junk:1999kv} for a Monte Carlo approach to
the calculation and to~\cite{clfft} for the analytic calculation
using Fast Fourier Transform (FFT) techniques. For completeness, we
introduce the basic approach below using the notation found
in~\cite{UW-confidence-VBF}.  For a counting experiment where we
expect, on average, $b$ background events and $s$ signal events, we
consider two hypotheses: the null (or background-only) hypothesis in
which the number of expected events, $n$, is described by a Poisson
distribution $P (n; b)$ and the alternate (or signal-plus-background)
hypothesis in which the number of expected events is described by a
Poisson distribution $P (n; s + b)$. Here the number of events serves
the purpose of a test statistic: a real number which quantities an
experiment.

It is possible to include a discriminating variable $x$ which has some
probability density function (pdf) for the background, $f_b(x)$, and
some pdf for the signal, $f_s(x)$, both normalized to unity. Given an
observation at $x$ we can construct the Likelihood Ratio $Q = (sf_s
(x) + bf_b (x))/bf_b (x)$. With several independent observations
$\{\hat{x}_i\}$ we can consider the combined likelihood ratio $Q = \prod
Q_i$ . It is possible, and in some sense optimal, to use $Q$ (or in
practice $q = \ln Q$) as a test statistic.

The computational challenge of using the log-likelihood ratio in
conjunction with a discriminating variable $x$ is the construction of
the log-likelihood ratio distribution for the background-only
hypothesis, $\rho_b(q)$, and for the signal-plus-background hypothesis
$\rho_{s+b}(q)$. In this case, there are not only the Poisson
fluctuations of the number of events, but also the continuously
varying discriminating variable $x$. In particular, for a single
background event the log-likelihood ratio distribution,
$\rho_{1,b}(q)$, must incorporate all possible values of $x$. From
these single event distributions we can build up the expected
log-likelihood ratio distribution by repeated convolution. This is
most effectively done by using a Fast Fourier Transform (FFT) where
convolution can be expressed as multiplication in the frequency domain
(denoted with a bar). In particular we arrive at:

\begin{eqnarray}\label{E:rho_def}
\overline{\rho_b(q)} &=& e^{b[\overline{\rho_{1,b}(q)}-1]} \hspace{.2in}{\rm and} \\ \nonumber
\overline{\rho_{s+b}(q)} &=& e^{(s+b)[\overline{\rho_{1,s+b}(q)}-1]}.
\end{eqnarray}
From the log-likelihood distribution of the two hypotheses we can
calculate a number of useful quantities. Given some experiment with an
observed log-likelihood ratio, $q^*$, we can calculate the
background-only confidence level, $CL_b$ :

\begin{equation}\label{E:clb}
CL_b (q^*) =\int_{q^*}^\infty \rho_b(q')dq'
\end{equation}
In the absence of an observation we can calculate the expected $CL_b$
given the signal-plus-background hypothesis is true. To do this
we first must find the median of the signal-plus-background
distribution $\overline{q}_{s+b}$.  From these we can calculate the
expected $CL_b$ by using Eq.~\ref{E:clb} evaluated at $q^* =\overline{q}_{s+b}$.

Finally, we can convert the expected background confidence level into
an expected Gaussian significance, $N\sigma$, by finding the value of
$N$ which satisfies
\begin{equation}
CL_b(\overline{q}_{s+b}) = \frac{1-{\rm erf}(N/\sqrt{2})}{2}.
\end{equation}
where ${\rm erf}(N) = (2/\pi) \int_0^N \exp( -y^2 )dy$ is a function
readily available in most numerical libraries.

\section{ Numerical Difficulties}\label{S:Numerical}
The methods described in the previous section have been applied to the
combined ATLAS Higgs effort with some caveats related to numerical
difficulties~\cite{UW-confidence-VBF}.  In particular, in the extreme
tails of $\rho_b(q)$, the probability density is dominated by
numerical noise. This numerical noise is an artifact of round-off
error in the double precision numbers used in the Fast Fourier
Transform\footnote{We use the FFTW library: http://www.fftw.org}. The
noise is on the order of $10^{-17}$ (for double precision floating
point numbers), which translates into a limit on the significance of
about $8\sigma$.  For particular values of the Higgs mass, ATLAS has
an expected significance well above $8\sigma$ with only $10~\rm
fb^{-1}$ of data. In order to produce significance values above the
$8\sigma$ limit, various extrapolation methods were used
in~\cite{UW-confidence-VBF}. We now introduce a definitive solution
to this problem based on arbitrary precision floating point numbers.

It should be made clear that the numerical precision problem is not
due to the fact that the $CL_b$ is so small that the evaluation of the
integral in Eq.~\ref{E:clb} cannot be treated with double precision
floating point numbers.  Instead, the numerical precision problem is
due to the many (approximately $2^{20}$) Fourier modes which must in
total produce a number very close to $0$. In order to rectify this
problem we have implemented the Fast Fourier Transform with the
arbitrary-precision floating point numbers provided in the CLN
library\footnote{CLN is available at
http://www.ginac.de}~\cite{GiNaC}.  One might protest that above
$5\sigma$ we are not interested in the precise value of the
significance and that this exercise is purely academic. We refer the
interested reader to Sections~\ref{S:Luminosity} \& ~\ref{S:Power} for
different summaries of an experiments discovery potential.

\begin{figure}
\epsfig{file=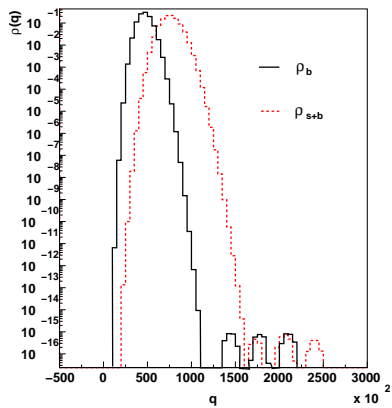, width=2in}
\caption{The distribution of the log-likelihood ratio $\rho(q)$ for
the null and alternate hypothesis (the axis labels refer to bins of
$q$, not $q$ itself). For $q > 10^5$ the distribution is contaminated
by numerical noise (see text for details).}
\label{fig:noise}
\end{figure}

\subsection{Extrapolation}\label{SS:Extrapolation}
\begin{figure}
\epsfig{file=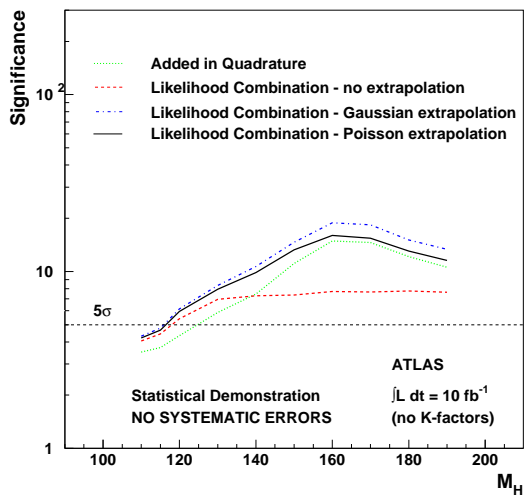, width=3in}
\caption{Comparison of the ATLAS Higgs combined significance obtained
from several approximate techniques. The (red) dashed line corresponds
to the unmodified likelihood ratio which can not produce significance
values above about $8\sigma$ (see text). This figure is meant to
demonstrate the different methods of combination and does not include
up-to-date results from the various Higgs analyses.}
\label{fig:extrap}
\end{figure}

While the arbitrary precision FFT approach is the definitive solution
to the problem of calculating very high expected significance, it is
also incredibly time consuming.  A much faster, approximate solution
is to approximate the $CL_b$ by fitting the $\rho_b$ distribution to a
functional form. The first method of extrapolation studied was a
simple Gaussian fit to the $\rho_b$ distribution. This method works
fairly well, but tends to overestimate the significance. The second
method we studied was based on a Poisson fit to the $\rho_b$ distribution. The Poisson distribution has the desirable properties that
it will have no probability below the hard limit $q \ge -s$ and that its
shape is more appropriate~\cite{UW-confidence-VBF}. Figure~\ref{fig:extrap}
compares these different extrapolation methods.

\section{Incorporating Systematic Uncertainty}\label{S:Systematics}

One encounters both philosophical and technical difficulties when one
tries to incorporate uncertainty on the predicted values $s$ and $b$
found in Eq.~\ref{E:rho_def}. In a Frequentist formalism the unknown $s$ and $b$ become
nuisance parameters. In a Bayesian formalism, $s$ and $b$ can be
marginalized by integration over their respective priors. At LEP the
practice was to smear $\rho_b$ and $\rho_{s+b}$ by integrating $s$ and
$b$ with a multivariate normal distribution as a prior. This smearing
technique is commonly referred to as the Cousins-Highland Technique,
and it is has some Bayesian aspects.

\subsection{A Purely Frequentist Technique}\label{SS:Freq}

At the PhysStat2003 conference a purely frequentist approach to
hypothesis testing with background uncertainty was
presented~\cite{Cranmer2003freq}. This method relies on the full
Neyman construction and uses a likelihood ratio similar to the profile
method as an ordering rule. In this formalism, a systematic
uncertainty at the level of 10\% has a much larger effect than when
treated with the Cousins-Highland technique.

\subsection{The Cousins­Highland Technique}\label{SS:CousinsHighland}
The Cousins-Highland formalism for including systematic errors on
the normalization of the signal and background is provided
in~\cite{Cousins:1992qz} and generalized in~\cite{Junk:1999kv, clfft}.
In particular, for a multivariate normal distribution\footnote{In
principle, any distribution could be used within this framework.}  as
a prior for the $n_i$ the distribution of the log-likelihood ratio is
given by:
\begin{eqnarray}\label{E:CousinsHighland}
\overline{\rho_{sys}(q)}=\int...\int
e^{\sum_i^{K} n_i[\overline{\rho_{1,i}(q)}-1]}
\left( \frac{1}{\sqrt{2\pi}}\right)^{K} \frac{1}{\sqrt{|S|}} \\ \nonumber
e^{\sum_i^{K} \sum_j^{K} -\frac{1}{2} 
(n_i - \langle n_i\rangle)S^{-1}_{ij}(n_j - \langle n_j \rangle)}
\prod_i dn_i
\end{eqnarray}
where $S_{ij} = \langle (n_i -\langle n_i \rangle)(n_j -\langle n_j
\rangle)\rangle$. Reference~\cite{clfft} provides an analytic
expression for the resulting log-likelihood ratio distribution
including a correlated error matrix; however, this equation was
obtained with an integration over negative numbers of expected events
and does not hold. Attempts to provide a closed form solution for the
positive semi-definite region require analytical continuation of the
error function over a wide range of the complex plane. Instead, a
numerical integration over the positive semi-definite region has been
adopted for our software packages.

\section{Discovery Luminosity}\label{S:Luminosity}
Because the calculation of expected significance is technically very
difficult at the LHC, other summaries of the discovery potential have
been explored. While these techniques are not new, it is important to
consider their pros and cons. One such alternate summary of the
discovery potential is based on the discovery luminosity". Define the
discovery luminosity, $L^*(m_H )$, to be the integrated luminosity
necessary for the expected significance to reach $5\sigma$. The
discovery luminosity is an informative quantity; however, it must be
interpreted with some care:

\begin{itemize}
\item Collecting an integrated luminosity equal to the nominal
discovery luminosity does not guarantee that a discovery will be
made. Instead, with $L^*(m_H )$ of data the median of $\rho_{s+b}$
will be at the $5\sigma$ level -- which corresponds to a 50\% chance of
discovery. See Section~\ref{S:Power} for more details.

\item In practice an analysis' cuts, systematic error, and signal and
background efficiencies are luminosity-dependent quantities. When we
calculate the discovery luminosity, we treat the analysis as constant.
\end{itemize}

\section{The Power of a $\bf 5\sigma$ Test}\label{S:Power}

The traditional quantity which is used to summarize an experiment's
discovery potential is the combined significance; however, as was
noted in Section~\ref{S:Numerical} this plot becomes very dificult to
make when the significance goes beyond about $8\sigma$. Furthermore,
the plot itself starts to loose relevance when the significance is far
above $5\sigma$. The discovery luminosity is another possible way of
illustrating an experiment's discovery potential, but it must be
interpreted with some care.  A third summary of an experiment's
discovery potential which is related to the probability of Type II
error: the {\it power}.  First, it should be noted that the expected
significance is a measure of separation between the {\it medians} of
the background-only and signal-plus-background hypotheses. Thus, when
we see the significance curve cross the $5\sigma$ line in
Fig.~\ref{fig:extrap} there is only a 50\% chance that we would
observe a $5\sigma$ effect if the Higgs does indeed exist at that
mass. In practice, we claim a discovery if the observed data exceeds
the $5\sigma$ critical region, and do not claim a discovery if it
doesn't. The meaning of the $5\sigma$ discovery threshold is a
convention which sets the probability of Type I error to be $2.85
\cdot 10^{-7}$ . With that in mind, the idea that the significance is
$20\sigma$ at $m_H = 160~{\rm GeV}$ is irrelevant. What is relevant is
the probability that we will claim discovery of the Higgs if it is
indeed there: that quantity is called the power. The power is defined
as $1-\beta$ where $\beta$ is the probability of Type II error: the
probability that we reject the signal-plus-background hypothesis when
it is true~\cite{Kendall}.

\begin{figure}
\epsfig{file=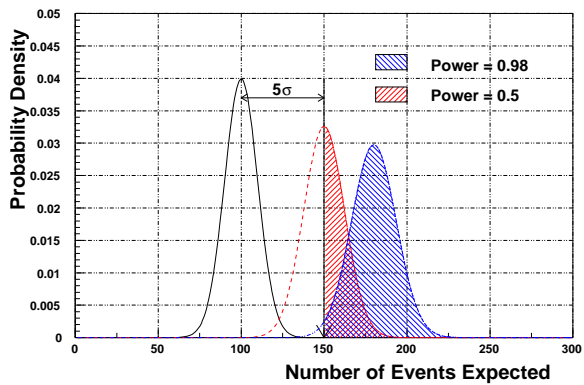, width=3in}
\caption{Examples of power for two different signal-plus-background
hypotheses with respect to a single background-only hypothesis with
100 expected events (black).}
\label{fig:power}
\end{figure}

Consider Figure~\ref{fig:power} with a background expectation of 100
events. The black vertical arrow denotes the $5\sigma$ discovery
threshold. The (red) dashed curve shows the distribution of the number
of expected events for a signal-plus-background hypothesis with 150
events. Normally, we would say the expected significance is $5\sigma$
for this hypothesis; however, we can see that only 50\% of the time we
would actually claim discovery.  The rightmost (blue) curve shows the
distribution of the number of expected events for a
signal-plus-background hypothesis with 180 events. Normally, we would
say the expected significance is $8\sigma$ for this hypothesis;
however, a more meaningful quantity -- the power -- is associated with
the probability we would claim discovery which is about 98\%. In
addition to the power being a germane quantity, it is much easier to
calculate.

\section{Conclusion}\label{S:Conclusions}

In conclusion, the migration of the statistical tool-set developed at
LEP to the LHC environment is not as straightforward as one might
expect. The first difficulties are computational and arise from the
combination of channels with many events and channels with few events
(these are easily solved). The next difficulties are numerical and
arise from the extremely high expected significance of the high-energy
frontier. These problems can be solved by brute force; or they can be
reinterpreted as conceptual problems, and solved by asking different
questions (i.e. power).  Lastly, there is a philosophical split
related to the Bayesian and Frequentist approach to uncertainty. At
the LHC, the choice of the formalism is no longer a second-order
effect, and this problem is not so easy to solve.

\begin{acknowledgments}
This work was supported by a graduate research fellowship from the
National Science Foundation and US Department of Energy Grant
DE-FG0295-ER40896.

\end{acknowledgments}

\bibliographystyle{unsrt}
\bibliography{vbf,bruce_cites,stats,nn,PhysicsGP}

\end{document}